# Spin fluctuation induced Weyl semimetal state in the paramagnetic phase of EuCd₂As₂

### Discovery of magnetic Weyl fermions: Dirac fermions split into pairs of Weyl fermions by slow magnetic fluctuations

J.-Z. Ma[1,3†], S. M. Nie[4†], C. J. Yi[2,5†], J. Jandke[1], T. Shang[1,3,6], M. Y. Yao[1], M. Naamneh[1], L. Q. Yan[2], Y. Sun[2,5,7], A. Chikina[1], V. N. Strocov[1], M. Medarde[6], M. Song[8,9], Y.-M. Xiong[8,10], G. Xu[11], W. Wulfhekel[12], J. Mesot[1,3,13], M. Reticcioli[14], C. Franchini[14,15], C. Mudry[16], M. Müller[16], Y. G. Shi[2,7*], T. Qian[2,7,17*], H. Ding[2,5,7,17], M. Shi[1]*

[1] *Swiss Light Source, Paul Scherrer Institute, CH-5232 Villigen PSI, Switzerland*

[2] *Beijing National Laboratory for Condensed Matter Physics and Institute of Physics, Chinese Academy of Sciences, Beijing 100190, China*

[3] *Institute of Condensed Matter Physics, École Polytechnique Fédérale de Lausanne, CH-10 15 Lausanne, Switzerland*

[4] *Department of Materials Science and Engineering, Stanford University, Stanford, CA 94305, United States*

[5] *School of Physics, University of Chinese Academy of Sciences, Beijing 100190, China*

[6] *Laboratory for Multiscale Materials Experiments, Paul Scherrer Institute, CH-5232 Villigen PSI, Switzerland*

[7] *Songshan Lake Materials Laboratory, Dongguan, Guangdong 523808, China*

[8] *Anhui Province Key Laboratory of Condensed Matter Physics at Extreme Conditions, High Magnetic Field Laboratory, Chinese Academy of Sciences, Hefei, Anhui 230031, China*

[9] *University of Science and Technology of China, Hefei, Anhui 230026, China*

[10] *Collaborative Innovation Center of Advanced Microstructures, Nanjing, 210093, China*

[11] *Wuhan National High Magnetic Field Center and School of Physics, Huazhong Science and Technology, Wuhan 430074, China*

[12] *Physikalisches Institut, Karlsruhe Institute of Technology, 76131 Karlsruhe, Germany*

[13] *Laboratory for Solid State Physics, ETH Zürich, CH-8093 Zürich, Switzerland*

[14] *Faculty of Physics, Center for Computational Materials Science, University of Vienna, A-1090 Vienna, Austria*

[15] *Dipartimento di Fisica e Astronomia, Università di Bologna, 40127 Bologna, Italy*

[16] *Condensed Matter Theory Group, Paul Scherrer Institute, CH-5232 Villigen PSI, Switzerland*

[17] *CAS Center for Excellence in Topological Quantum Computation, University of Chinese Academy of Sciences, Beijing 100190, China*

[†] These authors contributed to this work equally.

\* Corresponding authors: ming.shi@psi.ch, tqian@iphy.ac.cn, ygshi@iphy.ac.cn

Weyl fermions as emergent quasiparticles can arise in Weyl semimetals (WSMs) in which the energy bands are nondegenerate, resulting from inversion or time-reversal symmetry breaking. Nevertheless, experimental evidence for magnetically induced WSMs is scarce. Here, using photoemission spectroscopy, we observe that the degeneracy of Bloch bands is already lifted in the paramagnetic phase of EuCd$_2$As$_2$. We attribute this effect to the itinerant electrons experiencing quasi-static and quasi-long-range ferromagnetic fluctuations. Moreover, the spin-nondegenerate band structure harbors a pair of ideal Weyl nodes near the Fermi level. Hence, we show that long-range magnetic order and the spontaneous breaking of time-reversal symmetry are not an essential requirement for WSM states in centrosymmetric systems, and that WSM states can emerge in a wider range of condensed-matter systems than previously thought.

**INTRODUCTION**

In crystals, Kramers' theorem together with the combination of inversion ($P$) and time reversal ($T$) symmetries, protects the double degeneracy of fermionic energy bands. Dirac nodes can emerge at the gapless crossing of two doubly degenerate bands, near which the fermionic excitations are described by the massless 4×4 Dirac equation (*1-4*). The two-fold band degeneracy can be lifted when $P$ or $T$ symmetry is broken. The crossings of non-degenerate bands then leads to Weyl nodes which always occur in pairs. The electronic states with momenta close to a Weyl node are effectively governed by the 2×2 Weyl equation (*5*). While $P$-symmetry breaking is explicitly present in non-centrosymmetric systems, $T$ symmetry can be broken either explicitly by external magnetic fields or spontaneously through correlation effects. First-principle calculations have predicted a number of WSMs in non-centrosymmetric or magnetically ordered systems (*5-10*). From angle-resolved photoemission spectroscopy (ARPES) experiments, the Weyl nodes have been identified in several non-centrosymmetric systems, such as the TaAs family, (Mo,W)Te$_2$, LaAlGe, and TaIrTe$_4$ (*11-18*). By contrast, there is no well-defined spectroscopic evidence for Weyl nodes in magnetically ordered systems. Moreover, magneto-transport measurements have provided evidence for the chiral anomaly expected from Weyl fermions in TaAs, Na$_3$Bi, GdPtBi, Mn$_3$Sn and Co$_3$Sn$_2$S$_2$ (*19-23*). The Weyl nodes in Na$_3$Bi and GdPtBi are created by external magnetic fields.

Here, using ARPES we show that Weyl fermions emerge already in the paramagnetic (PM) phase of EuCd$_2$As$_2$. Together with measurements of transport, magnetic susceptibility, electron spin resonance (ESR), muon spin relaxation (μSR), and with first-principle calculations, we attribute the existence of these Weyl fermions to the effective breaking of *T* symmetry by ferromagnetic (FM) fluctuations on time and length scales that are long compared to the electronic ones, although the spontaneous *T* symmetry is preserved in the PM phase when considering dynamical statistics.

**RESULTS**

EuCd$_2$As$_2$ has a layered crystal structure with space group *P*-3*m*1 (No. 164). The Cd$_2$As$_2$ bilayers are separated by the triangular Eu layers (Fig. 1A). EuCd$_2$As$_2$ is an itinerant magnet with conduction electrons from the Cd and As orbitals. Magnetism originates from large local magnetic moments on the Eu ions. Previous studies revealed that the local Eu 4*f* moments form a long-range antiferromagnetic (AF) order with an A-type structure, i.e., FM *a-b* planes stacking antiferromagnetically along the *c* axis. This order sets in at the Néel temperature $T_N \sim 9.5$ K (*24-26*), at which both the resistivity $\rho(T)$ (Fig. 1A) and magnetic susceptibility $\chi(T)$ (Fig. 1C) show a peak.

Above $T_N$, the longitudinal-field (LF) μSR spectra show no obvious changes and no oscillations when different magnetic fields are applied in the PM phase of EuCd$_2$As$_2$ (Fig. 1G). In agreement with previous Mössbauer spectroscopy (*24*), the LF μSR results rule out any static magnetic order above $T_N$. Nevertheless, ESR measurements reveal that the resonance field ($H_{Res}$) starts to decrease around 100 K (Fig. 1E), indicating that (i) an effective internal magnetic field develops as the magnetic fluctuations slow down (*27*), and (ii) there is a relatively large scale for the magnetic interactions, most likely associated with strong coupling in the *a-b* plane. However, the $\chi^1$ (*T*) curves exhibit positive, but much lower Curie-Weiss temperatures $T_{CW}$, of the same order as $T_N$, both for magnetic fields (*H*) applied in the *a-b* plane or along the *c* axis (Fig. 1D), in agreement with previous measurements (*24,25*). The positivity of $T_{CW}$ suggests that magnetic fluctuations above $T_N$ are FM in nature, whereas its smallness as compared to the fluctuation temperature hints at the presence of competing interactions of either sign. Figure 1D shows that there is a crossing between the out-of-plane and in-plane susceptibility, the out-of-plane one being significantly larger at higher temperatures. Together this suggests predominantly FM fluctuations with out-of-plane magnetization deep in the PM phase.

Further evidence for the existence of a high characteristic temperature scale $T_F$ where fluctuations set in is found in transport. Figure 1F shows that an anomalous Hall effect (AHE) develops in the PM phase around 100 – 150 K already. At temperatures above 150 K, the Hall resistivity exhibits a simple linear dependence on the magnetic field. In contrast, it deviates from the linear behaviour when the temperature is below ~ 100 K. This is also an indication for the onset of quasi-static and quasi-long-range FM correlations at a fairly high $T_F$, since an AHE is typically related to either FM correlations or a nontrivial Berry curvature, associated with Weyl points. However, in the present contest, the latter requires an effective time-reversal symmetry breaking in the form of slow, large-scale FM fluctuations (*28-30*).

μSR is one of the most sensitive experimental methods to detect correlated fluctuating local magnetic fields (*31-34*) as one would expect from FM fluctuations in the present case. The ensuing exponential muon relaxation is indeed observed in zero field (ZF) and LF at different temperatures (Fig. 1G,H). When lowering the temperature, the dynamical muon relaxation rate $\lambda_{ZF}$, which is related to the fluctuating magnetic field, increases, starting at $\sim 100$ K. It then goes through a shoulder between 50 K and 100 K, and increases more steeply thereafter around $T_N$ (Fig. 1I).

Due to the layered crystal structure, the in-plane magnetic interactions are expected to be much stronger than the inter-layer magnetic interactions. This expectation is confirmed qualitatively by the following first principle calculations. When the measured lattice constants are imposed in the calculation, the ratio between the FM intra-plane and AFM inter-plane nearest-neighbour magnetic exchange couplings is estimated to be of order three in magnitude. We also find a substantial frustration between nearest- and next-nearest-neighbour magnetic exchange couplings. However, since the magnitude of these exchange couplings and even the sign of the inter-plane coupling change when the crystalline structure is allowed to relax during the calculation, a reliable quantitative prediction of their values is very difficult. The small experimental value of the ratio $T_N/T_F$ $\sim 0.1$ also suggests sizable frustration among the competing in-plane magnetic interactions, with the dominant one being FM and of order $T_F$. FM fluctuations in the plane may thus occur at temperatures below $T_F$. Once such quasi-static and quasi-long-range magnetic correlations with a magnetization pointing out-of-plane are established in the Eu planes by strong FM in-plane interactions, dipolar interactions will then further stabilize FM correlations out of the plane. True long-range FM order will however be prevented if, upon further cooling, the preferred magnetization changes from out-of-plane to in-plane orientation, in which case both dipolar and AF interlayer exchange interaction prefer an A-type AF magnetic structure. Such a change of in plane magnetization direction is exactly what is observed in Fig. 1D where the in-plane susceptibility starts dominating over the out-of-plane susceptibility below 17 K. This is consistent with recent resonant elastic X-ray scattering measurements which have confirmed that the magnetic moments lie in the *a-b* plane below $T_N$ ([26]).

If above $T_N$ a typical FM fluctuation is correlated over a characteristic linear size $\xi$ and over the characteristic time $\tau$ such that their inverses exceed the momentum and energy resolution needed to discern the lifting of the Bloch band degeneracy that is expected within the Born-Oppenheimer approximation due to the quasi-static FM, it will have a measurable effect on the itinerant electrons.

In ARPES experiments we observed the splitting of energy bands in the PM phase of $EuCd_2As_2$. In Fig. 2 we present the ARPES results acquired on cleaved (001) and (101) surfaces of $EuCd_2As_2$ crystals. Figure 2A,B shows a point-like Fermi surface (FS) at the $\Gamma$ point and cone-like band dispersions below the Fermi level ($E_F$). In Figs. 2C,D,H we display the FS in the $k_y - k_z$ plane and band dispersions along the $\Gamma A$ line which is normal to the (001) cleavage surface. The periodic appearance of the FS and the dispersive feature of the energy bands along $\Gamma A$ show that they are 3D bulk states. Figure 2E shows the band dispersions measured at 11 K along cut0 in Fig. 2A: There are flat bands located at $1 \sim 1.6$ eV below $E_F$ and several hole-like bands above or below the flat bands. While the flat bands arise from the Eu $4f$ orbitals, the hole-like bands mainly originate from the As $4p$ orbits. As the unit cell of $EuCd_2As_2$ contains two As atoms, there are in total six As $4p$ bands when both $T$ and $P$ symmetries are preserved. The interlayer coupling within the $Cd_2As_2$ bilayer results in three antibonding bands around $E_F$ and three bonding bands below the flat bands. Carefully examining the band dispersions it can be seen that a band splitting occurs near $E_F$, as indicated by arrows in Fig. 2F,G. In order to exclude the possibility that the band splitting results from surface effects or the broadening effect of the 3D bulk bands in the photoemission process, we have carried out soft X-ray ARPES measurements to increase the bulk sensitivity in the PM phase (Fig. 2H,I). The band splitting is still observed in the $k_x$-$k_y$ plane (Fig. 2I), providing evidence that it is an intrinsic effect of bulk states (*35*). We have also carried out ARPES measurements on the (101) surface. The APRES spectra in Fig. 2K,L were recorded with a photon energy of 88 eV. The corresponding momentum cut almost overlaps with the ΓA line near the Γ point (Fig. 2J). It should be mentioned that the band splitting is observed both in the raw ARPES spectra and in the curvature intensity plot, as indicated by arrows in Fig. 2K,L.

Usually, band splitting occurs when either $T$ or $P$ symmetry is broken. Since both $T$ and $P$ symmetries are preserved in the PM phase of EuCd$_2$As$_2$ when considering dynamical statistics over a large enough time scale, we attribute the observed band splitting to slow FM fluctuations with significant spatial correlations, both in-plane and out-of-plane. If the FM fluctuations are much slower than the relevant dynamical time for the itinerant electrons, a Born-Oppenheimer approximation by which magnetic fluctuations are treated as time-independent background fields for itinerant electrons is justified. A lower bound on the fluctuation time $\tau$ for the Born-Oppenheimer approximation to be justified is given by $\hbar/\Delta E_{split}$ with the magnetically induced band splitting $\Delta E_{split} \sim 0.1$ eV. This bound is well satisfied since the slow magnetic fluctuations could indeed be detected by μSR (with time resolution larger than picoseconds) which implies that $\tau \gg \hbar/\Delta E_{split}$. The effects of such slow fluctuations can thus be resolved in the spectral functions.

To provide further evidence that the band splitting is induced by spin fluctuations, we have performed ARPES experiments on related compounds in the same family and with the same crystal structure, one magnetic (EuCd$_2$Sb$_2$) and one non-magnetic (BaCd$_2$As$_2$). EuCd$_2$Sb$_2$ has very similar magnetic properties (such as an AFM phase transition at low temperature and spin fluctuations in the PM phase) as EuCd$_2$As$_2$, except a slightly lower Néel temperature $\sim 7.5$ K (*24*). The band splitting is very clear in EuCd$_2$Sb$_2$, as recorded with both a UV source and soft X-ray ARPES as shown in Fig. S1 of the Supplementary Material. It persists with increasing temperature up to 100 K above which the splitting cannot be well resolved (Fig. S2). In contrast, we did not observe any band splitting in $BaCd_2As_2$ (Fig. S3 in the Supplementary Material). We also confirm that the band structure above $T_N$ is very different from that in the low temperature AF phase, where a band folding occurs. We provide the associated experimental and numerical data in Fig. S4.

In order to study the band splitting in the PM phase of $EuCd_2As_2$, we have incorporated static FM patterns in our band structure calculations, assuming that fluctuations are much slower than the relevant dynamical time for the itinerant electrons. We carried out DFT + $U$ calculations with infinitude long-range FM order (assuming a large correlation length $\xi$) with the magnetic moments oriented in various directions. Here $U$ represents the Hubbard interaction among the Eu $4f$ orbitals. For simplicity, we start the discussion with the magnetic moments oriented along the $c$ axis. When $U = 0$, a number of flat bands associated with the Eu $4f$ orbitals appear near $E_F$ in Fig. S5A. Upon increasing $U$, the flat bands move downwards as shown in Fig. S5B-I. For $U = 5$ eV, the flat bands appear at $1 \sim 1.6$ eV below $E_F$ in Fig. 3A, in agreement with our ARPES results in Fig. 2E. In addition, the calculated band structures at $U = 5$ eV include several hole-like bands above and below the flat bands, which are consistent with the observation in Fig. 2. For comparison, the calculated band structure of $EuCd_2As_2$ in the absence of magnetic order is shown in Fig. S6. In this case the six As 4p bands collapse into three doubly degenerate hole-like bands protected by parity-time symmetry.

From our band structure calculations, one can see that the Cd $5s$ and As $4p$ states partially hybridize near $E_F$, for all values of $U$. However, when $U < 5$ eV, because of hybridization with the Eu 4f bands, the region near the Fermi level becomes rather tangled, as is clearly seen in Fig. S5. For $U \geq 5$ eV instead, an ideal band inversion develops near the Fermi level around the $\Gamma$ point. This results in an ideal band crossing of the Nth and (N+1)th bands at $k_z = \pm k_z^c$ on the $\Gamma$-A line in the presence of $C_{3z}$ symmetry, as shown in Fig. 3B. In any other $k_x$-$k_y$ planes with $k_z \neq \pm k_z^c$, the electronic structures are gapped at $E_F$, which makes it possible to define a Chern number $C$ for these planes. We found $C = -1$ for the planes with $|k_z| < k_z^c$ and $C = 0$ for the planes with $|k_z| > k_z^c$. Therefore, the band crossing points along the $\Gamma$-A line are topologically protected Weyl nodes. Band crossings resulting in Weyl points can also occur for the (N-1)th and Nth bands and/or the (N+1)th and (N+2)th bands. However, those Weyl nodes are typically farther away from $E_F$, and thus are less relevant for transport. Moreover, they are rather fragile due to the small band inversion, and can disappear through pairwise annihilation upon small changes of lattice parameters. Therefore, we focus on the lowest energy pair of Weyl nodes.

When considering FM fluctuations, for the theoretical modelling we make the following simplifying assumptions: (1) in the PM phase, at any given time, the system can be divided into FM correlation domains; (2) within each such domain the magnetic moments point in the same direction, whereby the orientation in different domains is random and uniformly distributed. (3) We assume the correlation length to be sufficiently large so that it does not introduce a significant uncertainty in k space. Then the FM fluctuations in the PM phase are captured by averaging the spectra of infinite large FM domains over the magnetic orientations. The resulting band structure along high-symmetry lines is shown in Fig. 3D. The average over directions preserves the band inversion around the $\Gamma$ point, and merely broadens the band structure by an amount $W_f$ (around $0.033 \pm 0.024$ Å$^{-1}$ along M-$\Gamma$-K, and around $0.024 \pm 0.012$ Å$^{-1}$ along $\Gamma$-A). $W_f$ is significantly smaller than the band splitting $W_s$ ($0.12 \pm 0.02$ Å$^{-1}$ along M-$\Gamma$-K, and $0.086 \pm 0.030$ Å$^{-1}$ along $\Gamma$-A) which can thus still be distinguished, as shown in Fig. 3E. The spin splitting $W_s$ observed in ARPES measurements is around $0.066$ Å$^{-1}$ along $k_z$=0 plane the and $0.033$ Å$^{-1}$ along $\Gamma$-A, respectively, and thus of the same order of magnitude. The agreement with predicted values is reasonable given that the above calculation neglects several sources of fluctuations and thus provides at best an upper bound for $W_s$. The effects of finite correlation lengths of the slow magnetic fluctuations on the visibility of the spin splitting are discussed in the supplementary text and Figs. S7, S8.

The variation of the location of the Weyl nodes upon changing the magnetic polarization direction in a correlation volume is moderate (also Table S1 in the supplementary material). For polarization along the $c$ axis, the $C_{3z}$ symmetry forces the Weyl nodes to lie on the high-symmetry line Γ-A. For different polarization they only deviate slightly from the Γ-A line. As illustrated in Fig. 3F, the pairs of parity-related Weyl nodes are confined in two small non-overlapping regions around the Γ-A line, the size of each region is less than 0.02 Å$^{-1}$, so the Weyl nodes in the PM phase of EuCd$_2$As$_2$ can be well detected despite fluctuations of the magnetic polarization direction.

According to the calculations, the separation of the Weyl nodes along the $k_z$ direction between is about 0.1 Å$^{-1}$, which is difficult to resolve in the ARPES spectra acquired from the measurements on the (001) surface because of the intrinsic low momentum resolution in the direction perpendicular to the surface. We thus carried out ARPES measurements on cleaved (101) surfaces, whose normal direction is along Γ-L (Fig. 1B). Figure 4A displays the FS intensity map measured from the cleaved (101) surface by varying the photon energy ($h\nu$). The momentum cut corresponding to $h\nu$ = 88 eV almost coincides with the ΓAΓ line (Fig. 4A). The three-dimensional (3D) ARPES intensity plot acquired at $h\nu \sim$ 88 eV exhibits two separate FS patches along the Γ-A direction (Fig. 4B,C) with cone-like band dispersions below $E_F$, in agreement with the calculations. In order to get an ARPES spectrum exactly on the Γ-A line, we collected a large set of ARPES data with photon energies in the vicinity of 88 eV. Figure 4D shows the band dispersions along Γ-A extracted from the photon energy dependent data, which has much better momentum resolution than that measured on the (001) surface along the same direction (Fig. 2D). The band splitting of the hole-like band can be identified both in the raw data and in the curvature intensity plot (Fig. 4D,E). In addition, we observed a shallow electron-like band (conduction band CB) near $E_F$ (Fig. 4F,H). As indicated in Fig. 4F,H,I, the electron- and hole-like bands, marked as CB and valence band (VB), respectively, cross $E_F$. As the electron band is also spin non-degenerate, the crossings of the Nth and (N+1)th bands lead to a pair of Weyl nodes around Γ-A, with $k_z \sim \pm 0.07$ Å$^{-1}$. To further confirm the crossings of conduction and valence bands, we carried out scanning tunneling microscopy/spectroscopy (STM/S) measurements on the (001) surface (Fig. 4J,K). The typical d$I$/d$U$ spectrum in Fig. 4K exhibits a "V" shape with a finite minimal intensity at $E_F$ and shows peaks both above and below $E_F$, which agree well with the inverted valence band top (VBT) and conduction band bottom (CBB), as marked in Fig. 4I. This is in agreement with the expected density of states of a Weyl-cone band structure.

In addition to the bulk Weyl nodes, a further important feature of Weyl semimetals are surface Fermi arcs. On the (101) cleaved surface, the two Weyl nodes are projected to different points in the surface BZ which should be connected by a surface Fermi arc. We indeed observe a signature of surface states in the photon energy dependent spectra as well as in the Fermi surface map. However, the Fermi arcs are rather sensitive to fluctuations of the polarization direction, which entails an intrinsic smearing of these surface states between the two Weyl points.

**CONCLUSION AND DISCUSSION**

We have shown that the Bloch bands of EuCd$_2$As$_2$ display topological attributes in the form of Weyl nodes that originate from the interplay between itinerant electrons and localized moments in the PM phase below the crossover temperature ~ 100 K. Far above 100 K, the Weyl nodes collapse into Dirac nodes as the typical lifetime or the spatial extent of the FM fluctuations become too short for the Kramers' degeneracy to be effectively broken. None of the probes used in this paper have direct access to the magnetic coupling of order 100 K that is responsible for the long-lived ferromagnetic fluctuations. The observation of Weyl nodes in $EuCd_2As_2$ is thus to be interpreted as the imprint on the Bloch bands of rather large competing magnetic interactions of the order of 100 K.

It should also be noted that the time scale for collecting ARPES spectra is much longer than that of the spin fluctuations. Therefore, ARPES data can be viewed as a statistical time average of dynamical results. The spin fluctuations do have an effect on the electronic structure, as they induce fluctuations on the Weyl nodes and the sign of the topological charge in the two Weyl node groups. These kinds of fluctuations, in Weyl node and charge sign, are sensitive to the magnetization, which can be tuned by an external magnetic field. The fact that Weyl nodes and their topological charge adjust to the orientation of the magnetization, which is itself tunable, could be promising for future spintronic applications.

A recent theoretical study has pointed out that a Dirac semimetal state or magnetic topological insulator state could coexist under certain conditions with long-range AF order (*36*). If this is also true for $EuCd_2As_2$ below $T_N$ the application of moderate magnetic fields could induce a metamagnetic transition to a FM state that would split the Kramers' degeneracy very substantially, even for a moderate field. We thus expect the phase diagram of $EuCd_2As_2$ as a function of temperature and magnetic field to be very rich.

**Materials and Methods**

Single crystals of $EuCd_2As_2$ were synthesized using Sn as flux. Starting materials of Eu (ingot, 99.9%, Alfa Aesar), Cd (grain 99.999%, Alfa Aesar), As (ingot, 99.999%, Alfa

Aesar) and excess Sn (grain 99.9969%, Alfa Aesar) were mixed and loaded in an alumina crucible at a molar ratio of 1:2:2:10. The operations were performed in a glove box filled with pure argon. Then, the crucible was sealed in a quartz tube under high vacuum. The tube was heated to 900°C and maintained for 20 hours before slowly cooling it to 500°C at a rate of 2°C /hour. Then the samples were separated from the Sn in a centrifuge.

ARPES measurements were performed at the SIS-HRPES beam line with a Scienta R4000 analyser and at the ADRESS beamline with a SPECS analyser of the Swiss Light Source (PSI), at the ARPES end-station of the Dreamline beamline at the Shanghai Synchrotron Radiation Facility (SSRF) and at the beamline UE112 PGM-2b-1^3 at BESSY Synchrotron. The energy and angular resolutions were set to 5~30 meV and 0.2°, respectively. The samples for ARPES measurements were cleaved *in situ* and measured in a temperature range between 2 K and 160 K in a vacuum better than $5\times10^{-11}$ Torr. The μSR measurements were carried out using the general-purpose spectrometers located at the πM3 beamline of the Swiss Muon Source of the PSI. STM measurements were carried out with a home-built Joule-Thomson STM (JT-STM) (37). The $EuCd_2As_2$ single crystals were cleaved in situ at T = 77 K. Measurements were performed at T = 2.8 - 14 K.

The DFT calculations were carried out by using the projector augmented wave method implemented in the Vienna ab initio simulation package (38,39). The cut-off energy for the plane wave expansion was 500 eV. The exchange-correlation functional was treated using the generalized gradient approximation (GGA) parametrized by Perdew, Burke, and Ernzerhof (40). SOC was taken into account self-consistently in the calculations. 10*10*5 k-point grids were used in the self-consistent simulations. The GGA+U method (41) was used to treat correlation effects in $EuCd_2As_2$. The s orbitals of Cd and the p orbitals of As were used to construct the maximally localized Wannier functions (MLWFs) (42), which were then used to calculate the Chern numbers.

To analyse the effects of the magnetic disorder with different FM clustering properties on the energy band structure of EuCd2As2, we carried out DFT+U calculations (U = 5 eV), including spin orbit coupling, by using a $4 \times 4 \times 1$ supercell (80 atoms). The magnetic moments of the 16 Eu atoms were constrained along arbitrary directions, exploring different arrangements. The band structure was unfolded onto the primitive cell Brillouin zone by adopting the unfolding method, as implemented in VASP (*43,44*).

**Acknowledgements**

We acknowledge E. Rienks, H. Luetkens, L.-Y. Kong, Y.-G. Zhong, H.-J. Liu, S.-Y Gao, X.-L Peng and Y.-B. Huang for the assistance during the measurements. We acknowledge Y. Goryunov, H.-M. Weng, L. Korosec, R. H. Ott, P. Hautle, Q. S. Wu, V. M. Katukuri and O. Yazyev for useful discussions. C.M. and M.M. thank B. Roessli for useful discussions. We acknowledge S. Blundell for sharing the μSR beamtime. **Funding**: This work was supported by the NCCR-MARVEL funded by the Swiss National Science Foundation, the Sino-Swiss Science and Technology Cooperation (Grant No. IZLCZ2-170075), the Ministry of Science and Technology of China (2016YFA0300600, 2016YFA0401000, 2017YFA0302901 and 2016YFA0300404), the National Natural Science Foundation of China (11622435, 11474340, 11474330, and 11674369), the Chinese Academy of Sciences (QYZDB-SSW-SLH043, XDB07000000, and XDPB08-1), and Beijing Municipal Science & Technology Commission (No. Z171100002017018).

**Author contributions**: M.S., T.Q. and J.Z.M supervised this project; J.Z.M. performed ARPES measurements with the assistance of M.Y.Y., A.C., and V.N.S; S.M.N. performed *ab initio* calculations with long range magnetic order background; M.R. and C.F. performed the band structure in the enlarged unit cell with different magnetic backgrounds, that differ in their average magnetization and clustering properties. C.J.Y. and Y.G.S. synthesized single crystals and performed resistance and magnetization measurements on single crystal $EuCd_2As_2$, $BaCd_2As_2$. M. Song. and Y.M.X. synthesized single crystal of $EuCd_2Sb_2$. T.S. and J.Z.M. performed AHE and μSR measurements. J.J., J.Z.M., and W.W. performed STM/STS measurements. L.Q.Y. and Y.S. performed ESR measurements. J.Z.M., T.Q., J.J., T.S., and M.S. analyzed the experimental data; J.Z.M.,

T.Q., M.S., and M.R. plotted the figures; J.Z.M., T.Q., M.S., M.M. and C.M. wrote the manuscript. All the authors discussed the results. **Competing interests**: The authors declare that they have no competing interests. **Data and materials availability**: All data needed to evaluate the conclusions in the paper are present in the paper and/or the Supplementary Materials. Materials and additional data related to this paper may be requested from the authors.

**Supplementary Materials**:

Materials and Methods

Figs. S1 to S8

Discussion on finitude correlation length disorder.

Tables. S1

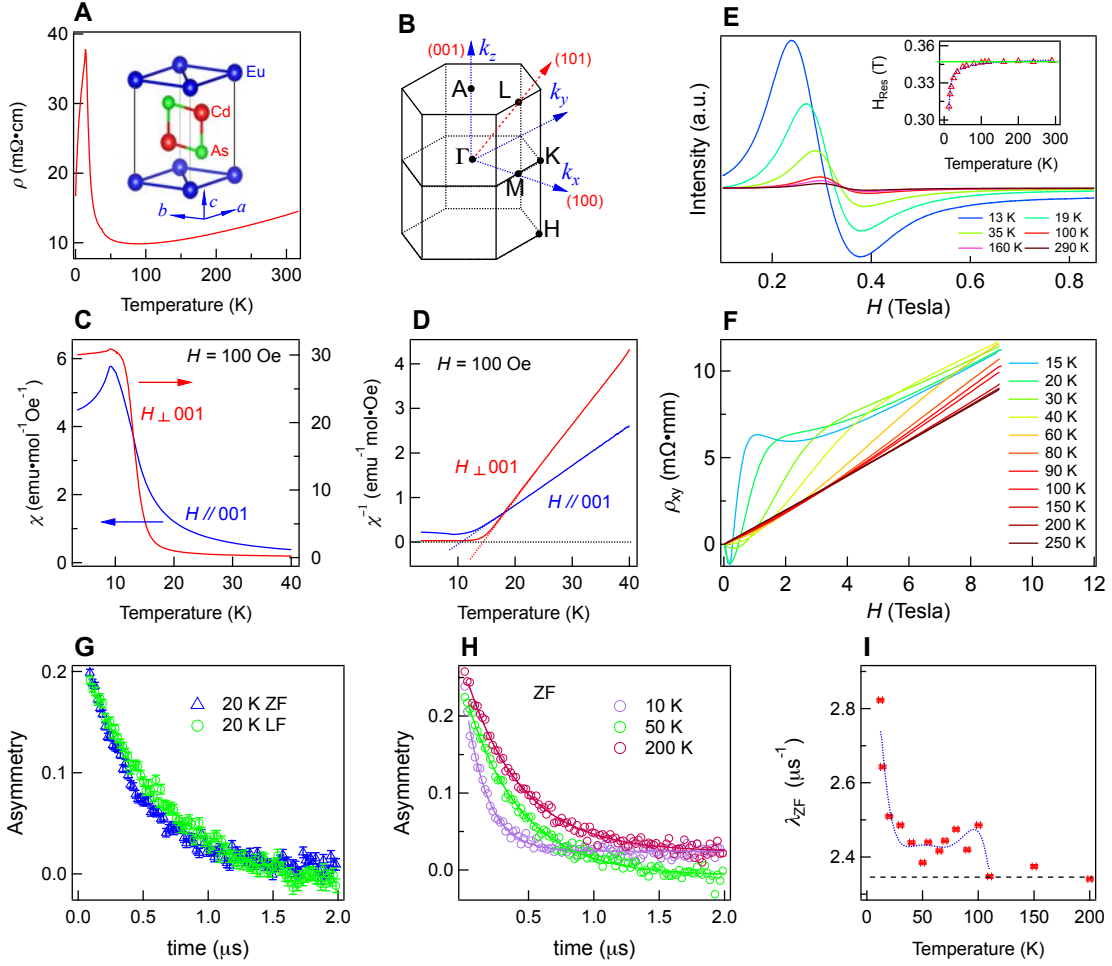

**Figure. 1. Slow FM fluctuations in the PM phase of EuCd₂As₂.** (**A**) Temperature dependence of the resistivity. The inset shows the crystal structure of EuCd₂As₂ in one unit cell. (**B**) 3D BZ with high-symmetry points and coordinate axes. The normal directions of cleaved (001) and (101) surfaces are also indicated. (**C**) Temperature dependence of the magnetic susceptibility with $H$ parallel and perpendicular to the $c$ axis, respectively. (**D**) Temperature dependence of the inverse susceptibility. (**E**) ESR spectra at various temperatures in the PM phase. The inset plots temperature dependence of the resonance field $H_{Res}$. (**F**) Magnetic-field dependence of the Hall resistivity at various temperatures under in-plane magnetic fields. (**G**) μSR spectra at 20 K in zero field (ZF) and longitudinal field (LF) 7000 Oe, respectively. (**H**) μSR spectra at three representative temperatures. (**I**) Temperature dependence of the dynamic muon relaxation rate $\lambda_{ZF}$.

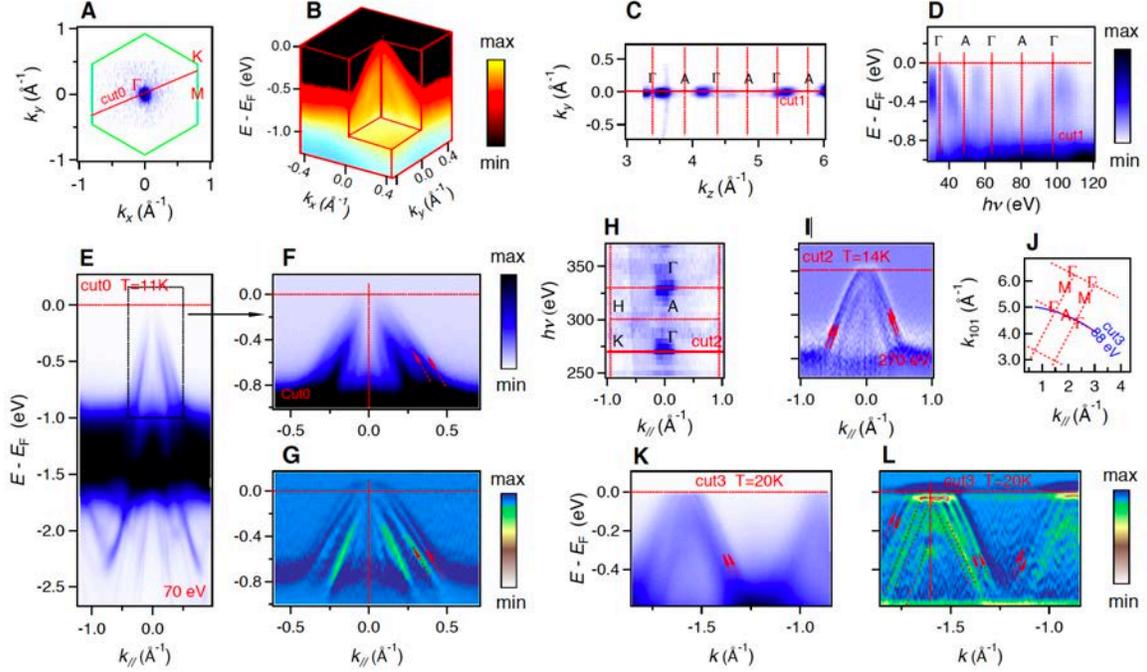

**Figure. 2. Band splitting in the PM phase of EuCd$_2$As$_2$.** The ARPES data in (A-J) were taken from the cleaved (001) surface, the data in l-m were collected from the cleaved (101) surface. (**A**) ARPES intensity map at $E_F$ acquired with photon energy 70ev, showing a point like FS in the $k_x$-$k_y$ plane. (**B**) 3D intensity plot of the ARPES data collected with $hv$ = 70 eV, showing cone-shape dispersions in the $k_x$-$k_y$ plane. (**C**) Intensity plot of the ARPES data at $E_F$ collected with $hv$ varying from 30 to 130 eV, showing the FSs in the $k_y$-$k_z$ plane. (**D**) ARPES spectrum image along cut1 in the Γ-A direction as shown in **c**. (**E**) A detailed view of the band structure along cut0 as shown in **a** collected at $T$ = 11 K. (**F, G**) Zoom-in of the raw ARPES spectra in the box shown in (E) and the corresponding 2D curvature intensity plot, respectively. (**H**) Intensity plot of the ARPES data at $E_F$ collected with $hv$ varying from 250 to 360 eV, showing the FSs in the $k_y$-$k_z$ plane. (**I**) Curvature intensity plot of the ARPES data along Γ-K (cut2 in (H)) collected at $T$ = 14 K with $hv$ = 270 eV. (**J**) The profile of Brillouin Zones (BZs) in the plane along normal direction passing through Γ, A, M points and perpendicular to the (101) surface. (**K, L**) The ARPES spectrum and corresponding 2D curvature intensity plot along cut3 in (J). The arrows indicate the band splitting.

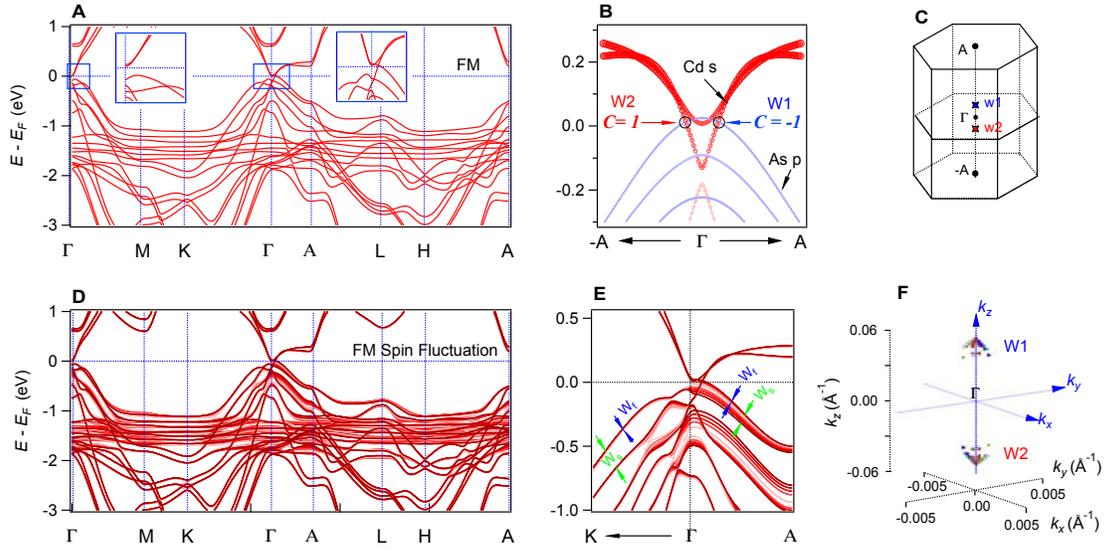

**Figure. 3. The band structures in the FM phase and PM phase with FM fluctuations.**
(**A**) The band structure along high-symmetry lines calculated by using DFT + U with $U =$ 5 eV. The magnetic moments oriented in the (001) direction. The insets are the zoom-in of band dispersions in the vicinity of the $E_F$ around the Γ point. (**B**) In the near $E_F$ region the bands along Γ-A with blue and red colors scaling the components of As 4$p$ and Cd 5$s$ orbitals, respectively. The blue circles indicate different Weyl points. (**C**) Locations of Weyl points (W1 and W2) in the 3D BZ. (**D**) The same as that in (A), but the electronic structure is a superposition of the energy bands calculated with the magnetic moments in all possible directions to simulate the FM fluctuation. (**E**) The zoom-in of the superposed electronic structure along K-Γ-A in the vicinity of $E_F$, indicating the broadening effect induced by spin fluctuation. $W_s$ and $W_f$ are width of spin splitting and the band broadening, induced by the FM fluctuation, respectively. (**F**) The distribution of Weyl points (W1 and W2), calculated with the magnetic moments oriented in all kinds of typical directions.

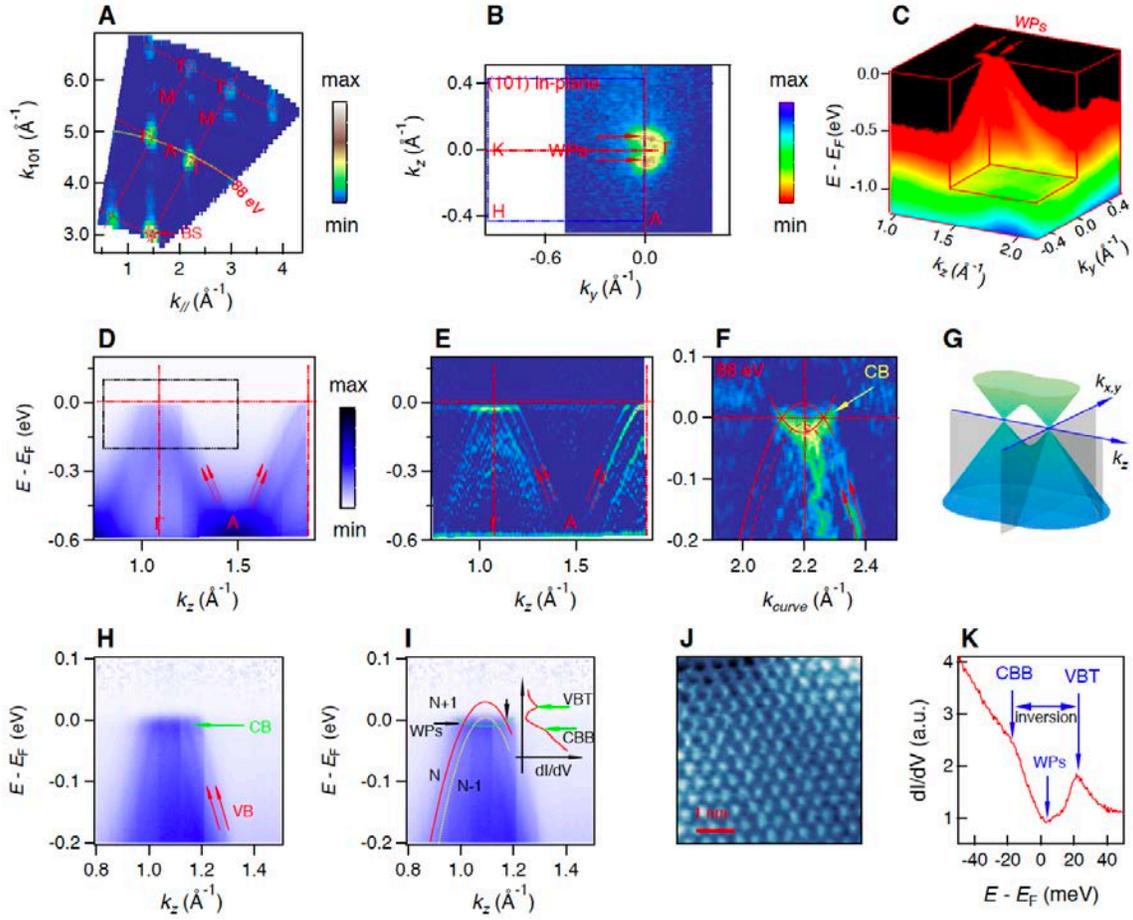

**Figure. 4. The observation of Weyl cones.** Unless other mentioned, the spectra are collected from the cleaved (101) surface using $hv$ = 88 eV, the corresponding momentum cut is indicated with the yellow curve in (A), which almost overlaps with the Γ-A- Γ line. (**A**) The FS map in the $k_x$-$k_z$ plane, collected with $hv$ in the range of 30 - 160 eV. (**B**) The FS map in the $k_y$-$k_z$ plane. (**C**) The 3D ARPES intensity to show two point-like FSs on Γ-A and the cone-shape dispersions in the $k_y$-$k_z$ plane. (**D,E**) The ARPES spectrum and its corresponding curvature intensity plot along Γ-A. To obtain the ARPES spectrum on a straight line located exactly on the Γ-A line, a number of high-resolution ARPES data were collected with photon energies in the vicinity of 88 eV. The arrows indicate the band splitting. (**F**) The 2D curvature intensity plot of the ARPES spectrum along the yellow curve in (A). An electron conduct band with band bottom below $E_F$ is clearly visible. (**G**) Schematic of the 3D Weyl-cone band structure in the $k_x$-$k_z$ ($k_y$-$k_z$) plane. (**H, I**) Zoom-in of the dashing-line box in (D), the spectrum was divided by Fermi-Dirac function. CB, VB, VBT and CBB stand for conduction band, valence band, the top of valence band and the bottom of conduct band, respectively. Black arrows point to the Weyl points. The tunneling differential conductance (dI/dV) curve from STS measurements is plotted for comparison. (**J**) STM constant current topographic image obtained from (001) surface of $EuCd_2As_2$. (**K**) The dI/dV spectrum recorded at $T = 11$ K on the (001) surface. In this figure the data in A-C and F (D-E, H-I) were collected from sample #1 (sample #2). We note that sample #1 is slightly less hole-doped than sample #2, which makes the electron band easier to be explored in sample #1 in (F).

# Supplementary information of

# Spin fluctuation induced Weyl semimetal state in the paramagnetic phase of EuCd₂As₂

J.-Z. Ma[1,3†], S. M. Nie[4†], C. J. Yi[2,5†], J. Jandke[1], T. Shang[1,3,6], M. Y. Yao[1], M. Naamneh[1], L. Q. Yan[2], Y. Sun[2,5,7], A. Chikina[1], V. N. Strocov[1], M. Medarde[6], M. Song[8,9], Y.-M. Xiong[8,10], G. Xu[11], W. Wulfhekel[12], J. Mesot[1,3,13], M. Reticcioli[14], C. Franchini[14,15], C. Mudry[16], M. Müller[16], Y. G. Shi[2,7*], T. Qian[2,7,17*], H. Ding[2,5,7,17], M. Shi[1*]

Corresponding authors: ming.shi@psi.ch, tqian@iphy.ac.cn, ygshi@iphy.ac.cn

**This PDF file includes:**

Figs. S1 to S8.

Discussion on finitude correlation length disorder.

Tables. S1.

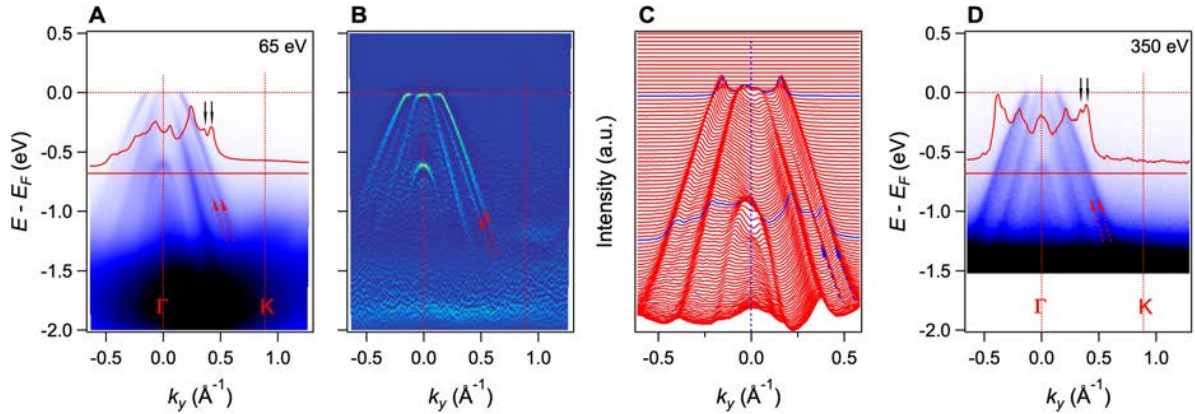

**Figure. S1. Band splitting in EuCd₂Sb₂. (A,B)** The ARPES spectrum and its curvature intensity plot along Γ-K. The data was collected at 15K with photon energy $h\nu$ = 65 eV. The red arrows indicate the split bands. The red curve in (A) is the momentum

distribution curve (MDC) along the constant energy cut indicated by the straight line below the MDC. The double peak pointed out by the vertical arrows results from the splitting of the outer band, as indicated by the red arrows. (**C**) A stack of MDCs extracted at different energies from the spectrum in (A). For clarity the MDC curves are offset successively. (**D**) The ARPES spectrum along Γ-K, the data was acquired using soft X-rays with $hv$ = 350 eV. The band splitting was observed using both VUV-light and soft X-rays, which provide compelling evidence that band splitting is an intrinsic feature of the bulk states rather than a $k_z$ broadening effect.

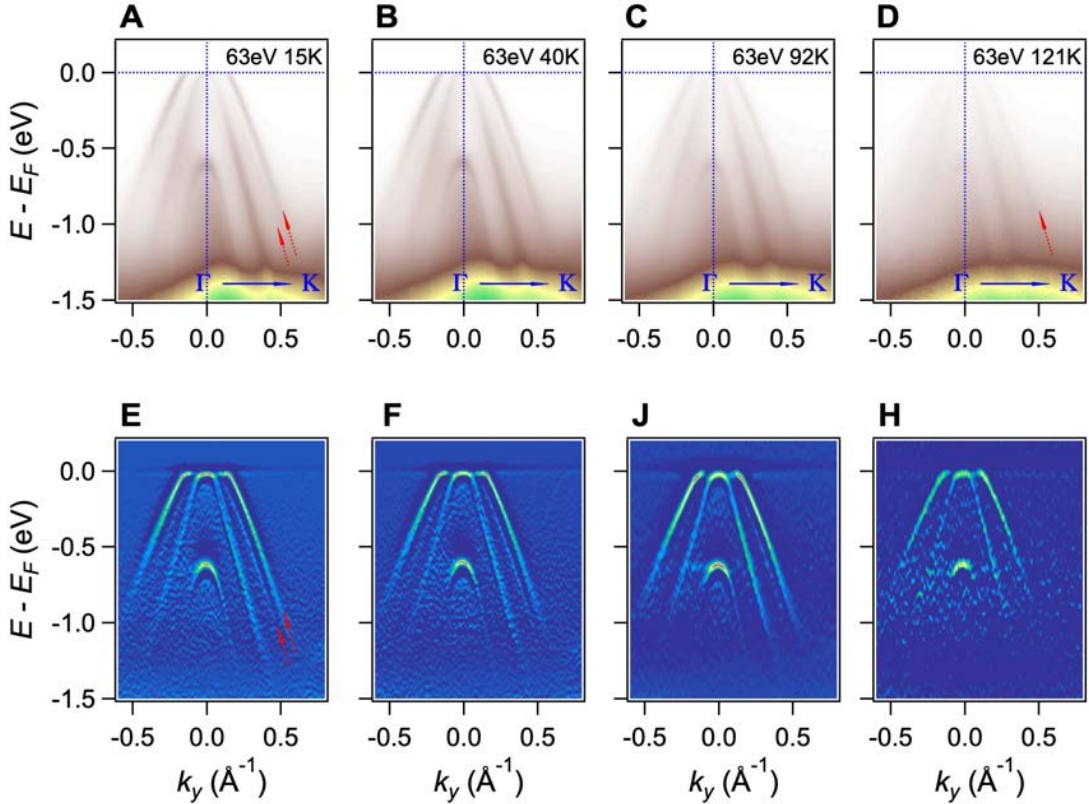

**Figure. S2. Temperature effects on the band splitting in EuCd$_2$Sb$_2$.** (**A** to **D**) The ARPES spectrum along Γ-K, as a function of temperature from 15 K to 121 K, recorded with $hv$ = 63 eV. The two arrows in (A) indicate the band splitting at low temperature.

Upon increasing the temperature, the split bands become blurred and are hard to distinguish above 100 K. (**E** to **H**) The curvature intensity plots of the spectra in (A to D).

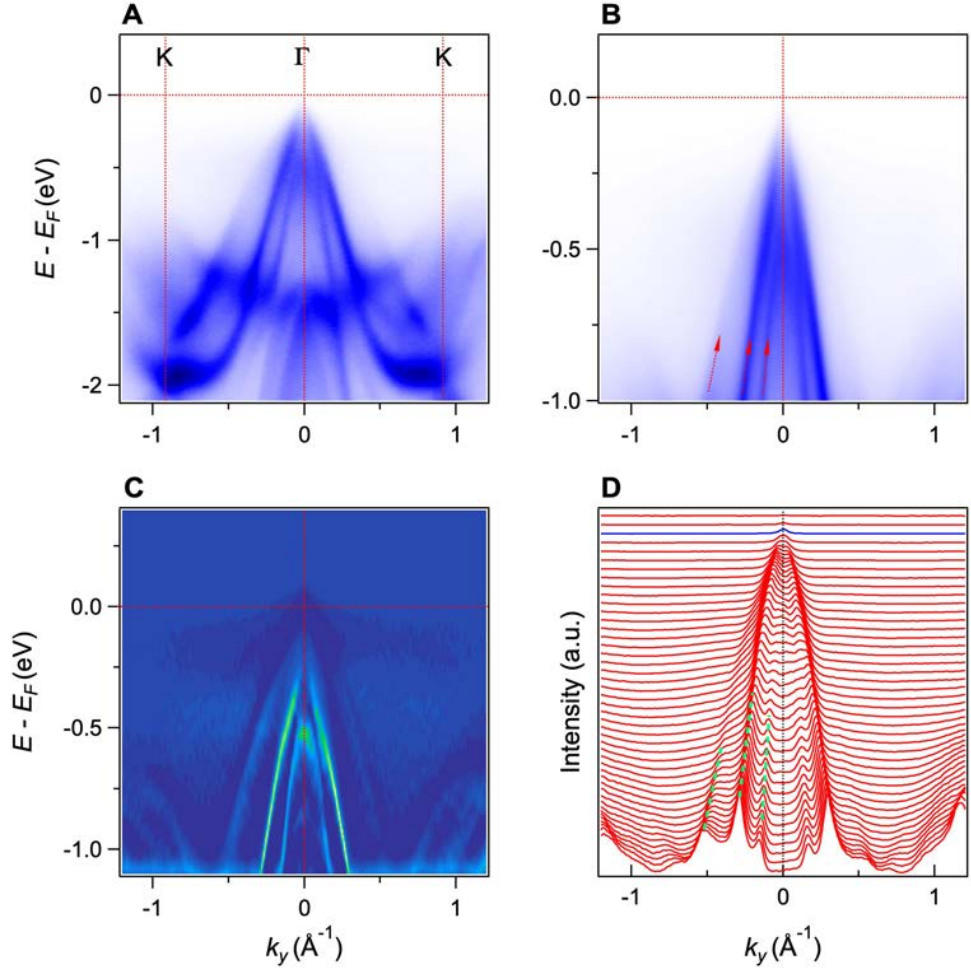

**Figure. S3. Electronic structure of BaCd₂As₂.** (**A,B**) The ARPES spectrum along K-Γ-K at two different energy scales. (**C**) Curvature intensity plot of the spectrum in (B). (**D**) A stack of MDCs extracted from the spectrum in (A) at different energies. For clarity the MDC curves are increasingly offset with increasing energy. The blue line indicates the Fermi level. No band splitting is observed in this non-magnetic compound, neither in the raw ARPES data nor in the curvature intensity plots or the MDC plots as indicated by the green dashed lines that trace the three doubly degenerate bands.

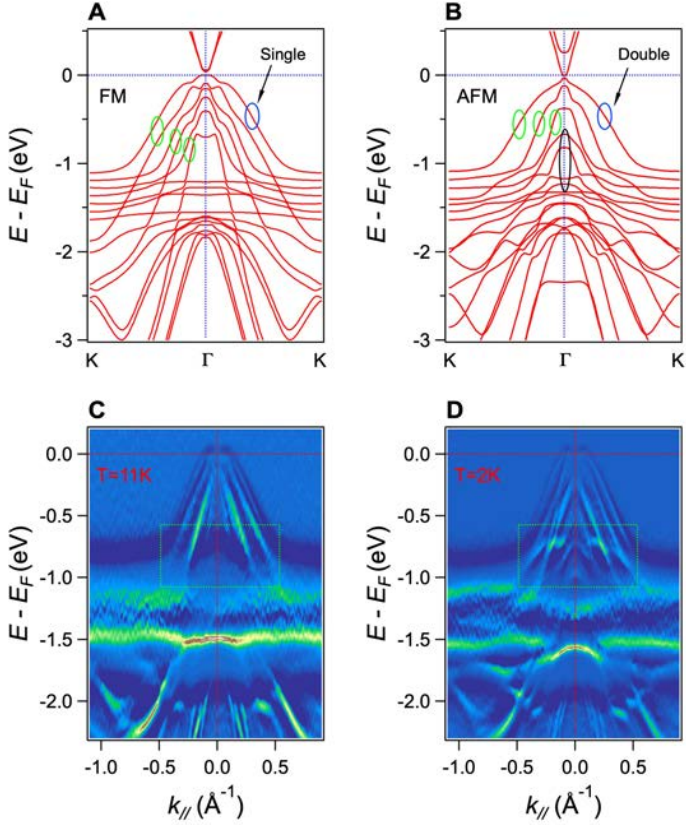

Fig. S4. Comparison of band structures below and above Néel temperature. (**A**) Calculated band structure along K-Γ-K in the FM phase. (**B**) Calculated band structure along K-Γ-K in the AFM phase. (**C**) Curvature intensity of band structure along cut0 as shown in Fig. 2A recorded at T=11K. (**D**) The same as C but recorded at T = 2 K. For the FM calculation shown in A, there are six singly degenerate As 4p bands near the Fermi level with their band top above -0.6 eV. These six bands organize into three spin-split pairs, as indicated with green ellipses. In the AFM calculation, shown in B, the six As 4p bands collapse into three bands indicated with green ellipses. Moreover, since the unit cell in the AFM phase is doubled along the z direction with respect to the FM phase, the BZ of the AFM is folded. The folding from the A point to the Γ point results in additional bands at Γ, as indicated by the black ellipse in B. It should be noted that, even though time reversal symmetry ($T$) is broken in the AFM phase, the double degeneracy of the bands is protected by the combination of parity symmetry ($P$), time reversal symmetry ($T$) and translation symmetry ($L$) by one unit along the z-axis. More details can be found in (*36*). The ARPES data in the PM phase agrees well with the FM band calculation, as we discuss in the main text. As we cool below the Néel temperature, we observe three hole-like bands near the Fermi level with their band top above -0.4 eV. Moreover, a few shallow bands appear between -0.5 to -1.0 eV (region indicated by the green box) which we interpret as the folded bands discussed in the AFM calculation. These folded shallow bands cannot be observed in the PM phase above $T_N$.

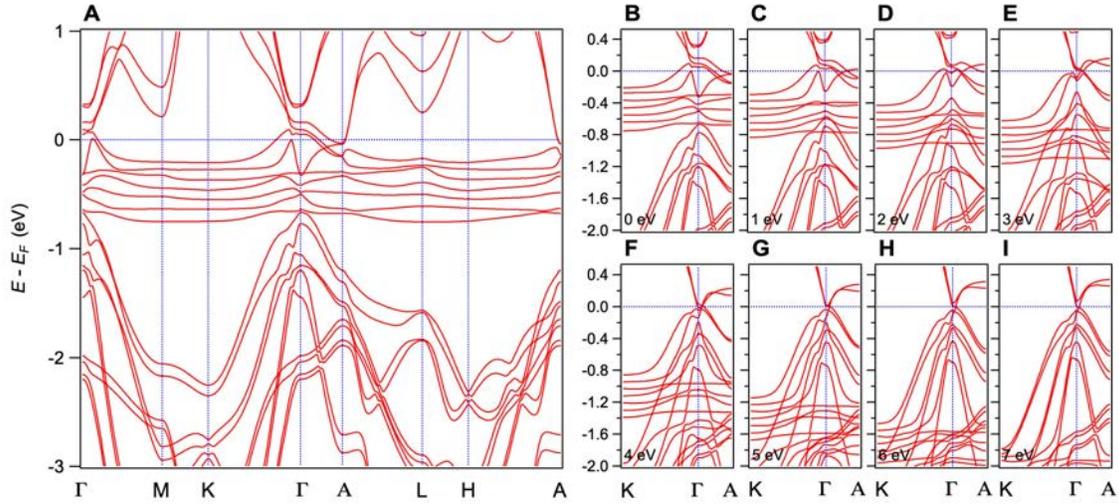

**Figure. S5.** **Calculated band structures of EuCd₂As₂ with magnetic moments oriented along the c-axis, as a function of onsite Coulomb interaction U.** (**A**) Band dispersions along high-symmetry lines, calculated with U = 0 eV. (**B** to **I**) bands along K-Γ-A with U values from 0 to 7 eV.

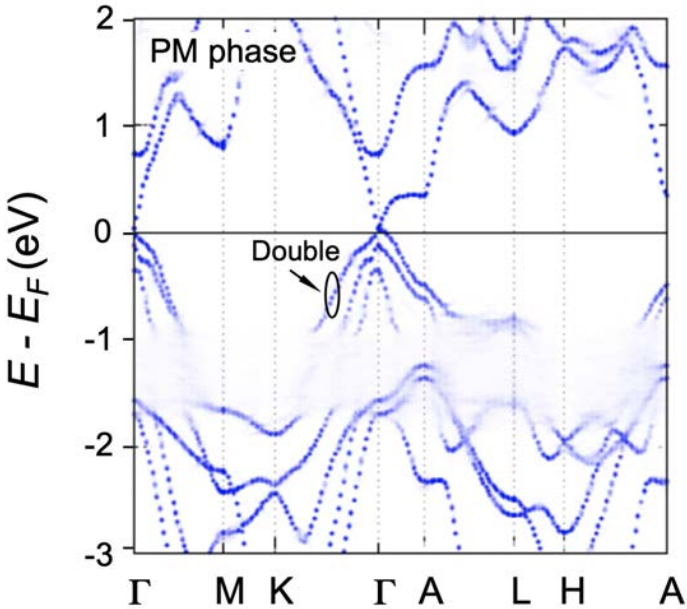

Fig. S6. Calculated band structure of EuCd₂As₂ along high symmetry lines, deeply within the PM phase. Here, we set random directions for the Eu magnetic moments in a 4*4*1 enlarged unit cell (with a negligible average magnetization) in order to describe slow, but uncorrelated magnetic fluctuations. The results are unfolded to the single-formula-unit primitive cell. The six previously spin-split As 4p bands essentially collapse into three hole-like bands, as indicated in the figure for one of the three bands, and no discernible spin splitting is observed.

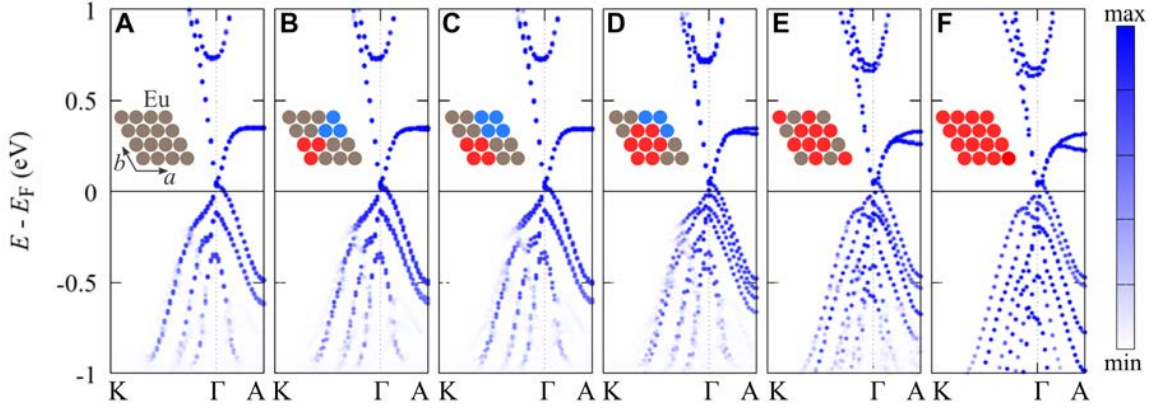

Fig. S7. Here, we consider an enlarged unit cell with different magnetic backgrounds, that differ in their average magnetization and clustering properties (or correlations): (**A**) Completely random spin configuration. (**B**) to (**E**) Structures containing small ferromagnetic clusters. (**F**) Ferromagnetic configuration. For each magnetic background we calculate the band structure along KΓA. We observe a spin splitting, which increases with the size of the included ferromagnetic clusters, and/or magnetization. However, to compare with the setting of an ARPES experiment, one should generate all possible magnetic structures, weighted with the Boltzmann weight of the associated spin patterns and average the resulting spectra. As explained in the main text, we expect this average to wash out the spin splitting of the bands, unless the correlation length is larger than the inverse of $W_s$. More details please see following discussion part.

**Spin-splitting and band-broadening by static disorder with finite correlation length**

Here we analyze in more detail, how quasi-static magnetic fluctuations with different spatial correlations affect the electronic spectrum, going beyond the infinite range correlations discussed in the main text. We treat the magnetic fluctuations arising from the magnetic moments of the Eu 4f orbitals as follows. In the spirit of the Born-Oppenheimer approximation, we model the magnetic fluctuations by a statistical ensemble of random energies $B(x)$, such as Zeeman energies, say, with the Gaussian distribution

$$\overline{B(x)} = 0, \tag{1a}$$

$$\overline{B(x)B(x')} = B_*^1 e^{-|x-x'|/\xi}. \tag{1b}$$

The disorder thus depends on two parameters, the standard deviation $B_\star$ and the correlation length $\xi$, which is the correlation length below which the microscopic Eu spins are ferromagnetically aligned in three-dimensional space. For simplicity we assume correlations to be isotropic. Note that $\xi$ is bounded from below by the lattice spacing $a$. As a function of temperature $T$, $\xi$ reaches a maximum above the antiferromagnetic ordering temperature $T_N$ and reaches the lattice spacing $a$ upon approaching $T_N$ from above, as well as $T \to \infty$. A ferromagnetic alignment of spins introduces a spin splitting of order $B_\star$. In momentum space the dispersion curves are split by $\Delta k_S$ (denoted $W_S$ in the main text), where

$$\Delta k_s = \frac{B_*}{\hbar v_F}. \tag{1c}$$

For a gas of itinerant electrons for which spin is a good quantum number the Fermi energy $\varepsilon_F$ is related to the Fermi wave number $k_F$ and the Fermi velocity $v_F$ through

$$\varepsilon_F = \hbar v_F k_F. \tag{2}$$

For a good metal, $1/k_F$ is of the order of the lattice spacing $a$.

Perturbing this spin-degenerate electron gas with random static magnetic fields results in the mean-free path $l$, i.e., the scattering length within the Born approximation. Its dependence on the dimensionless numbers $B_\star/\varepsilon_F$ and $k_F\,\xi \gg 1$ is

$$l \sim \left(\frac{\varepsilon_F}{B_*}\right)^2 \frac{1}{k_F\xi}\frac{1}{k_F}. \tag{3}$$

(The factor $k_F\,\xi$ that enters the Born scattering time arises from having a multiplicative gain in forward or backward scattering by the number $\xi^d$ of parallel spins in the correlated disordered region, divided by the factor $\xi^{d-1}$ accounting for the fact that scattering is confined to small angles $\sim 1/k_F\xi$. In other words, the problem reduces essentially to 1-dimensional ray optics in the limit $\xi \gg 1/k_F$. For 1d problems the decrease of the mean free path $\sim 1/\xi$ is indeed well established.) According to Eq. (3), the mean-free path decreases as $B_*$ or $\xi$ increase, as long as the random magnetic energies can be treated perturbatively. If we define the length scale

$$\xi_* \equiv \frac{\hbar v_F}{B_*} \equiv \frac{1}{\Delta k_X} \equiv \frac{1}{W_s}, \tag{4a}$$

we may combine Eqs. (2) and (3) into

$$l \sim \left(\frac{\xi_*}{\xi}\right)\xi_*. \tag{4b}$$

The spin splitting $B_\star$ produced by a spatially homogeneous magnetic configuration should be compared to the energy uncertainty $\hbar v_F/\xi$ arising from confining a ballistic electron at the Fermi energy in a box of linear size $\xi$. The spin splitting becomes only important if

$$B_* \gtrsim \frac{\hbar v_F}{\xi}, \tag{5a}$$

i.e.,

$$\xi \gtrsim \xi_*. \tag{5b}$$

In the regime

$$a \leq \xi \ll \xi_*, \tag{6a}$$

the effects of the static random magnetic energies is to give a small uncertainty to the single-particle dispersion of the unperturbed electron gas, i.e., the $k$-dependence of the electronic spectral function is a narrow Lorentzian of width $1/l$ instead of a delta function (see left inset in Fig. S8). No spin splitting can be resolved within this narrow Lorentzian. Indeed, the electronic motion cannot be confined to a correlation volume, and thus averages over the random magnetic field. However, once

$$\xi \sim \xi_*, \tag{6b}$$

 the effect of the static random magnetic energies on the single-particle dispersion of the spin-degenerate electron gas is non-perturbative. The k-dependence of the electronic spectral function still has a broad single peak as a function of $k$, but with a width of order $1/\xi_\star = \Delta k_S$ (see right middle inset in Fig. S8). In the regime

$$\xi \gg \xi_* \,, \tag{6c}$$

because the electrons travel in sufficiently large regions in which the magnetic field is uniform, the energy gain $B_*$ resulting from splitting the Kramers' degeneracy in the bulk of these correlated regions is much larger than the energy uncertainty due to scatterings at their (random) boundaries. The $k$-dependence of the electronic spectral function becomes a double Lorentzian with the two peak maxima separated by $\Delta k_S \equiv W_S$, which is larger than the width $1/\xi$ of either one of the two peaks (see right inset in Fig. S8). Hence, the splitting of the Kramers' degenerate bands now can be resolved. The resolution increases with increasing $\xi$. In particular, the mean free path is not anymore given by Eq. (3), but simply tracks $\xi$. The qualitative dependence on $\xi$ of the mean-free path in the regimes (6a) and (6c) is shown in Fig. S8 assuming that the random magnetic energies can be treated perturbatively around the limiting cases $\xi \to a$ (no spin splitting of the bands) and $\xi \to \infty$ (two spin split bands).

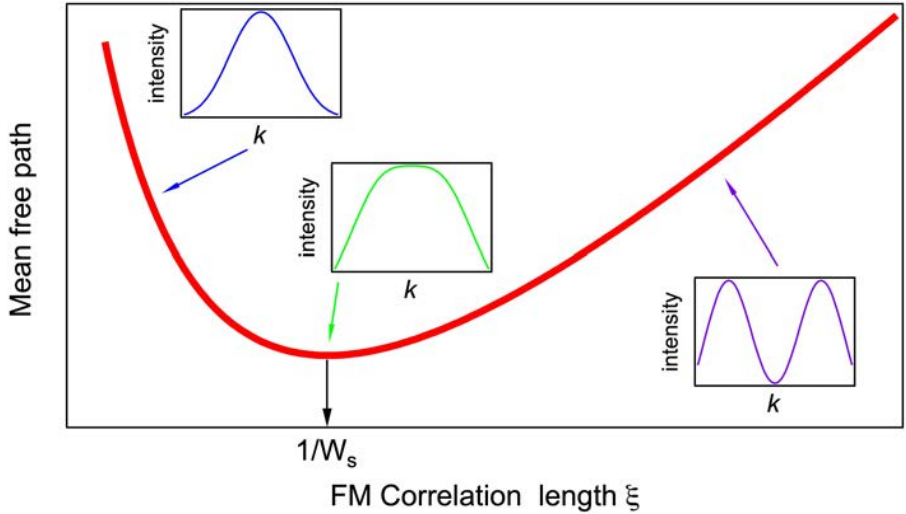

Fig. S8. The mean free path as a function of the FM correlation length.

**Table. S1**. Positions of the Weyl points in EuCd$_2$As$_2$ depending on the spin orientation. $\theta$ is the angle between the spin orientation and the $c$-axis. $\varphi$ is the angle between the projection of the spin orientation onto the $a$-$b$ plane and the $b$ axis.

| $\theta$ | $\varphi = 0°$, Weyl point position: $k_x$ (Å$^{-1}$), $k_y$ (Å$^{-1}$), $k_z$ (Å$^{-1}$)) |
|---|---|
| 0° | (0.0000, 0.0000, 0.0438), (0.0000, 0.0000, -0.0438) |
| 30° | (0.0001, 0.0001, 0.0423), (0.0000, -0.0001, -0.0423) |
| 45° | (0.0000, 0.00015, 0.0414), (0.0000, -0.00015, -0.0414) |
| 60° | (0.0000, 0.00018, 0.0401), (0.0000, -0.00018, -0.0401) |
| 75° | (0.0000, 0.0002, 0.0386), (0.0000, -0.0002, -0.0386) |
| 90° | (-0.00008, -0.00015, 0.0335), (0.00008, 0.00015, -0.0335) |
| $\theta$ | $\varphi = 90°$ |
| 30° | (0.00016, 0.00004, 0.0423), (-0.00016, -0.00004, -0.0423) |
| 45° | (0.00026, 0.00009, 0.0408), (-0.00026, -0.00009, -0.0408) |
| 60° | (0.00037, 0.00016, 0.0394), (-0.00037, -0.00016, -0.0394) |
| 75° | (0.00048, 0.00023, 0.0374), (-0.00048, -0.00023, -0.0374) |
| 90° | (0.00014, 0.0007, 0.0316), (-0.00014, -0.0007, -0.0316) |
| $\theta$ | $\varphi = 45°$ |
| 30° | (0.00016, 0.00011, 0.0423), (-0.00016, -0.000011, -0.0423) |
| 45° | (0.00029, 0.00018, 0.0408), (-0.00029, -0.00018, -0.0408) |
| 60° | (0.00043, 0.00024, 0.0394), (-0.00043, -0.00024, -0.0394) |
| 75° | (0.00058, 0.00033, 0.0374), (-0.00058, -0.00033, -0.0374) |
| 90° | (0.00011, 0.0002, 0.0330), (-0.00011, -0.0002, -0.0330) |